\documentclass[twocolumn,secnumarabic,amssymb, nobibnotes, aps, prapplied]{revtex4-2}

\usepackage[pdftex]{graphicx}
\usepackage{amsmath}
\usepackage{amssymb}
\usepackage{chngcntr}
\usepackage{float}
\usepackage{ifpdf}
\usepackage{subcaption}
\usepackage{caption}
\usepackage{xcolor}
\begin{document}

\title{Absorption and scattering by a temporally switched lossy layer: Going beyond the Rozanov bound}

\author{Chen Firestein and Amir Shlivinski}
\email{amirshli@bgu.ac.il}
\affiliation{School of Electrical and Computer Engineering, Ben-Gurion University of the Negev, Beer Sheva, 84105, Israel}

\author{Yakir Hadad}
\email{hadady@eng.tau.ac.il}
\affiliation{School of Electrical Engineering, Tel-Aviv University, Ramat-Aviv, Tel-Aviv, Israel, 69978}

\begin{abstract}
In this paper we study the electromagnetic scattering, absorption, and performance bounds for short time modulated pulses that impinge on a time-varying lossy layer that is sandwiched between vacuum and a perfect electric conductor. The electric characteristics of the  layer, namely, the conductivity, permittivity, and permeability are assumed to change \emph{abruptly or gradually} in time.
We demonstrate numerically that a time-varying absorbing layer that undergoes temporal switching of its \emph{permittivity and conductance} can absorb the power of a modulated, ultra-wideband, as well as a quasi-monochromatic, pulsed wave \emph{beyond} what is dictated by the time invariant Rozanov bound \emph{when integrating over the whole frequency spectrum}.
We suggest and simulate a practical metamaterial realization that is constructed as a three-dimensional array of resistor loaded dipole. By switching only the dipole's load resistance, desired effective media properties are obtained.  Furthermore, we show that Rozanov's bound can be bypassed with abrupt and a more practical gradual, soft, switching thus overcoming some possible causality issue in abrupt switching.

\end{abstract}
\maketitle

\section{Introduction}
Wave absorbers find important applications in various wave based  systems, in acoustics and  electromagnetics, where they are used  for the minimization of scattering and emission from certain domains, as for example radar absorbers or in electromagnetic compatibility applications \cite{Sievenpiper2018,luo2019subwavelength,landy2008perfect,li2014metamaterial,ruck1970radar,park2001photonic,qu2021conceptual, Assouar2016}. The  typical absorber  is based on a proper spatial arrangement of linear and time-invariant (LTI) lossy materials. For LTI  absorbers that consist of a lossy layer backed by an ideal conductor (PEC), Rozanov \cite{rozanov2000ultimate} has developed a fundamental tradeoff between the absorber thickness and the absorption bandwidth. However, since this bound, as several other bounds in wave theory, has been developed under the LTI assumption, it does not necessarily apply when non-LTI e.g. linear time varying (LTV) wave systems are considered.

In fact, time-varying wave devices have been shown in recent years to allow an additional degree of freedom in compare to conventional LTI designs and therefore in many cases they provide a potential to achieve better performances \cite{shlivinski2018beyond,li2019beyond,guo2020improving,li2020temporal,solis2020generalization,chen2013broadening}.
Interestingly, the time-variation is also considered as a means for wave manipulations leading to additional peculiar functionalities
\cite{caloz2019spacetime1,caloz2019spacetime,elnaggar2020modelling,taravati2019generalized,deck2019uniform,hadad2015space, chamanara2019simultaneous,sedeh2020time,taravati2020full,barati2020topological,sedeh2020adaptive,del2020reconfigurable,shi2016dynamic, pacheco2020antireflection,kazemi2019exceptionaln,kazemi2020ultra,rouhi2020exceptional}.
As spatial variations, also temporal variations can take different forms, such as periodic modulation and temporal discontinuity. The former is better suited for narrowband applications \cite{li2019beyond, hadad2020antenna, loghmannia2019broadband}, whereas the latter fits better with pulsed wave applications. Specifically, it has been used in \cite{shlivinski2018beyond} to achieve a better impedance matching between TL and reactive loads for short-time pulses, by \cite{pacheco2020antireflection} for the purpose of anti-reflection coating, and by \cite{li2020temporal} to improve the reflection bandwidth of a Dallenbach screen. Furthermore, the problem of reflection and transmission, as well as energy balance,  in the presence of abrupt and gradual temporal variation of the guiding medium were explored in  \cite{xiao2011spectral,xiao2014reflection,tan2020energy,hadad2020soft,koutserimpas2020electromagnetic,lurie2017energy,lurie2016energy,mirmoosa2018unlimited,kiasat2018temporal,koutserimpas2018electromagnetic}.
Scattering and radiation by time-modulated small obstacles that can be described using the dipole approximation have been explored in \cite{hadad2019space,mirmoosa2020instantaneous,ptitcyn2019time,mirmoosa2020dipole}. Such description may be helpful for the development of homogenization methodologies for time varying composite media and may be used to practically emulate effective bulk metamaterials with various time-varying characteristics, as regained e.g. in our work.

In this paper we study the electromagnetic wave scattering and absorption of short time pulses that impinge on a time-varying lossy dielectric layer that is sandwiched between half-space of vacuum and a perfect electric conductor. The electric characteristics of the layer, namely, its conductivity, permittivity, and permeability are assumed to change \emph{abruptly or gradually} in time. To make the analysis as general as possible we study the equivalent transmission line (TL) problem of the actual electromagnetic system.
In accordance with Rozanov's bound, the incident pulse is assumed to impinge normally which enforces a TEM mode \cite{collin2007foundations,rao2004elements}.
 We solve a temporally switched TL problem in the complex frequency plane $s$ and by performing an inverse Laplace transform the time domain voltage is obtained for any point along the TL. We use this approach to augment the Rozanov bound for ultra-wideband and quasi-monochromatic short-time pulsed signals followed by a demonstration of how it may be bypassed using a temporally-switched layered wave absorber.
While our time-domain absorption analysis has a broader applicability, specifically to bypass the LTI bound we only apply temporal switching of the layer's permittivity and conductivity which simplifies practical realizations. To that end we suggest a metamaterial structure that is composed of an array of resistor loaded dipoles. By changing the \emph{resistance only} we obtain the required effective media properties. Moreover, we demonstrate that by using a realistic gradual soft switching instead of an abrupt one, the absorbtion of the time variant system is still better than the Rozanov bound, even if the switching duration is several times larger than the pulse temporal width.

\section{Description of the Problem}
\begin{figure}[H]
  \center
  \includegraphics[height=6cm, width=1\columnwidth]{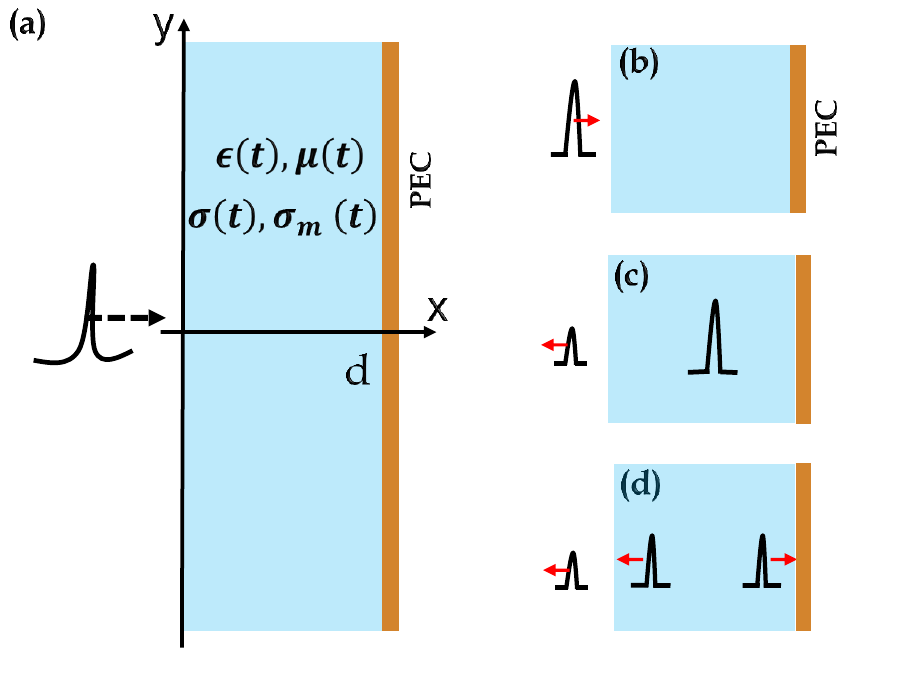}
  \caption{ Problem illustration; (a) A short pulse is propagating towards a radar absorber that is composed of magneto-dielectric layer(blue) attached to a Perfect Electric Conductor (PEC) surface(brown). The figure illustrates the $(x,y)$ plane, which is the plane of incidence.\\
  Temporal description of the system: (b) At $t=t_{0}$ the incident wavefront impinges the boundary. (c) At time $t=t_{s}$ the wave is completely contained within the layer and an abrupt switching is preformed. (d) For times $t>t_{s}$ the electric and magnetic properties may lead to dispersive reflected and a transmitted waves.}
  \label{Fig.1.}
\end{figure}

An electromagnetic pulsed wave with electric and magnetic fields, $\vec{E}(\mathbf{r},t)$ and $\vec{H}(\mathbf{r},t)$,  propagates in vacuum in the normal direction towards an absorbing layer. At $t=t_0$ the pulsed wave impinges a dielectric layer with permittivity, permeability, conductivity and magnetic conductivity ($\epsilon_1,\mu_1,\sigma_1,\sigma_{m1}$). The layer thickness is $d$, and it is sandwitched between vacuum and an ideal electric conductor. See  Fig.~\ref{Fig.1.}(a) for illustration.
With respect to the plane of incidence, the wave is of a Transverse Electromagnetic (TEM) mode which enables the use of an equivalent TL model.
Following standard textbook formulation such \cite{collin2007foundations,rao2004elements}, the TL model is readily available.
%
%
We assume that initially the layer is lossless and can be described by the following TL parameters $\{L_{1},C_{1}\}$,[$\{R_{1}=0,G_{1}=0\}$], where $\{L_{1},C_{1}\}$ [$\{R_{1},G_{1}\}$] are the per unit length inductance and capacitance [series resistance and parallel admittance].
Due to the spatial discontinuity at $x=0$ some of the impinging wave will be back reflected upon incidence (see Fig.~\ref{Fig.1.}c). This reflection is non-dispersive, i.e., the transmitted and back reflected pulses do not change their shape. It has a simple reflection coefficients of $\Gamma_{s}=(Z_1-Z_0)/(Z_1+Z_0)$, where $Z_{0}=\sqrt{L_0/C_0}$ is the characteristic TL impedance modeling the host medium (vacuum), and $Z_1=\sqrt{L_1/C_1}$ is the characteristic TL impedance of the layer.

At time $t_{s}>t_{0}$, while the pulse is completely contained within the layer (see Fig.~\ref{Fig.1.}.(c)-(d)), the TLs' parameters are abruptly switched into a new set of parameters $(\epsilon_2,\mu_2,\sigma_2,\sigma_{m2})$ which translates into $\{L_2,C_2,R_2,G_2\}$ in the TL model. Thus forming a \emph{temporal boundary} separating between two TL problems, with initial conditions set at $t=t_{s}^{+}$ along the TL. In general, this switching process  ignites two waves that are counter-propagating inside the layer \cite{comm2}. %
These two waves maintain the general pulsed shape, however their temporal width may be different than that of the pulse at $t<t_s$, giving rise to pulse compression or expansion. In association with that, the switching process may absorb or pump energy into the wave system \cite{shlivinski2018beyond,xiao2014reflection}.

\section{Mathematical Formulation}
In this paper we strive to calculate the overall reflection by the temporally switched layer. This includes, of course, the reflection by the spatial discontinuity at $x=0$, but also it includes the reflection due to the temporal discontinuity resulting from the switching process, and the energy change caused by it.
In order to solve this scattering and absorption problem, in the next section, we develop a time-domain Green's function formalism for the complete analysis of the fields at times $t>t_{s}$. The wave solution of the pulsed waveform along the layer for $0<x<d$ and $t>t_{s}$ is obtained by a superposition integration of the Green's function weighted by an initial time source at $t=t_{s}^{+}$ that results from fluxes continuity.

\subsection{The TL model and its formal solution}
\begin{figure}[H]
  \center
  \includegraphics[width=0.8\columnwidth]{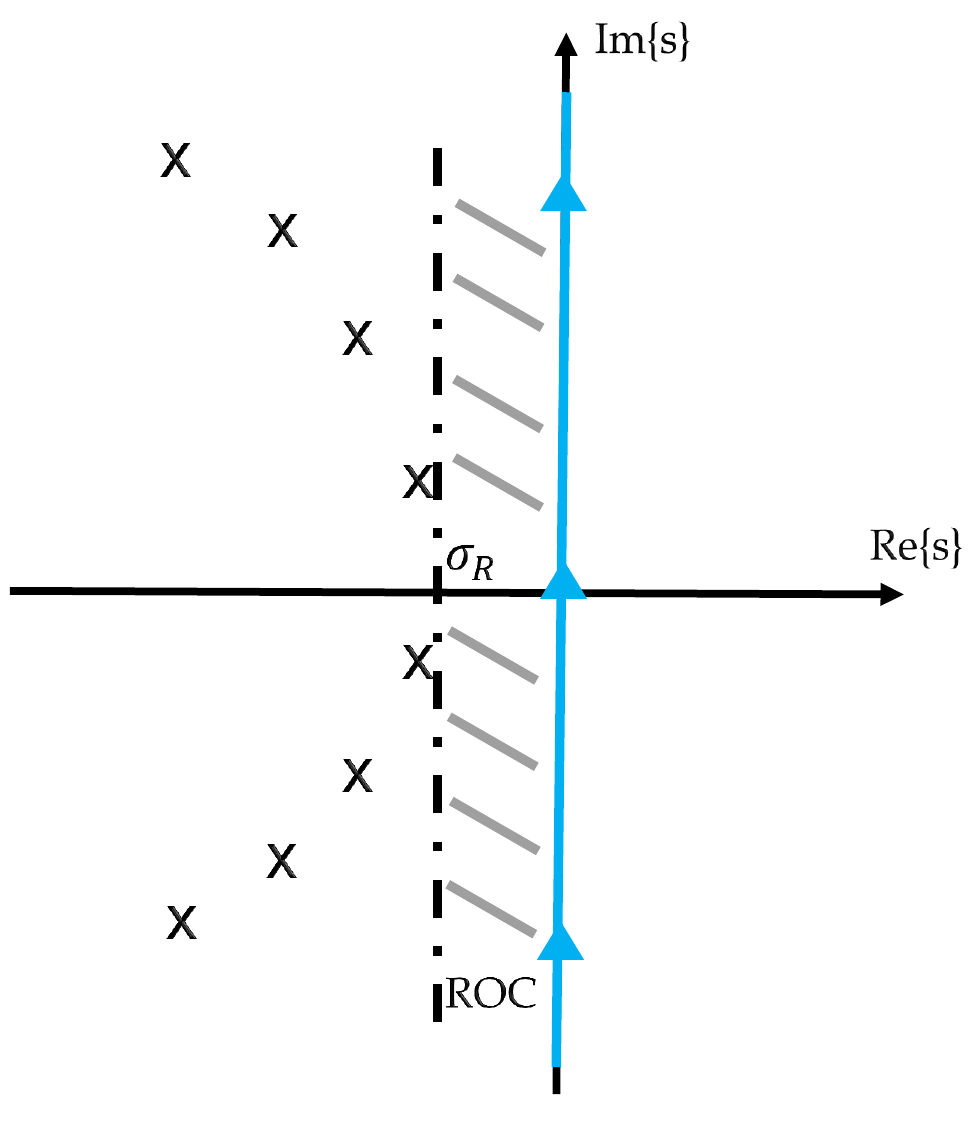}
  \caption{Time domain voltage is evaluated by performing an integration over imaginary axis which states for an inverse Laplace transformation. $\sigma_{R}$ is the real part of the rightmost singularity of $V(x,s)$.}
  \label{Fig.3.}
\end{figure}

We reduce the general plane wave scattering problem into a $1$D-TL problem that is governed by the TLs equations
\begin{subequations}\label{2}
    \begin{eqnarray}
    &&\frac{\partial V(x,t)}{\partial x} +L \frac{\partial I(x,t)}{\partial t}+RI(x,t)=0 \\[1ex] \label{1.a}
    &&\frac{\partial I(x,t)}{\partial x} +C \frac{\partial V(x,t)}{\partial t}+GV(x,t)=0 \label{1.b},
    \end{eqnarray}
\end{subequations}
where $V(x,t)$ and $I(x,t)$ denote the voltage and current along the line, respectively, and are related to the field in each domain.
As a first step,  we transform the time-domain system in Eq.~(\ref{2}) into the complex frequency (Laplace) domain
\begin{subequations}\label{3}
    \begin{eqnarray}
    \label{3a}
    &&\frac{\partial V(x,s)}{\partial x} +(sL+R)I(x,s) -LI_0(x)=0\\[1ex]
    \label{3b}
    &&\frac{\partial I(x,s)}{\partial x} +(sC+G)V(x,s) -CV_0(x)=0.
    \end{eqnarray}
\end{subequations}
where $V_0(x)$ and $I_0(x)$ are the initial voltage and current at $t=t_{s}^{+}$ respectively, and $s$ denotes the, complex frequency spectral variable. Taking the $x$ derivative of \eqref{3a} followed by the substitution of \eqref{3b}, gives the Helmholtz equation for a lossy medium,
\begin{subequations}\label{4}
  \begin{eqnarray}
  \label{4a}
  &&\frac{\partial^2 V(x,s)}{\partial x^2} -(sL+R)(sC+G)V(x,s)=f_{s,v}(x,s)\hspace{0.2cm}\\[1ex]
  \label{4b}
  &&f_{s,v}(x,s)=L\frac{\partial I_0(x)}{\partial x}-C(sL+R)V_0(x),
  \end{eqnarray}
\end{subequations}
with impedance boundary conditions at $x=0$, and short circuit due to the ideal conductor at $x=d$. We define the spectral Green function for the voltage, $g_v(x,x',s)$, such that the solution for the spectral voltage $V(x,s)$ can be expressed as the following superposition integral
\begin{equation}\label{7}
  V(x,s)=\int_{-\infty}^{\infty} g_v(x,x',s)f_{s,v}(x',s)dx'.
\end{equation}
From this definition, clearly, $g_v(x,x',s)$ is a solution of
\begin{equation}\label{5}
  \frac{\partial^2 g_v(x,x',s)}{\partial x^2} -\gamma ^2g_v(x,x',s)=\delta (x-x'),
\end{equation}
where $\gamma=\sqrt{(sL+R)(sC+G)}$, with $\mbox{Re\{$\gamma$ \}}>0$, and $\delta$ denotes the Dirac delta function. Once the voltage $V(x,s)$ is solved, the time domain voltage is obtained by the inverse Laplace transform,
\begin{equation}\label{8}
V(x,t)=\frac{1}{2\pi j} \int_{\kappa-j\infty}^{\kappa+j\infty}V(x,s)e^{st}ds,
\end{equation}
where $-\sigma_R<\kappa<0$ in the region of convergence (ROC) as shown in Fig.~\ref{Fig.3.}. In light of its realizability, our time-varying electromagnetic system is causal. Moreover, since it involves a single temporal switching, as opposed to periodic variation, it is necessarily stable. As a result,  all the pole singularities of $V(x,s)$  must lie in the left side of the complex frequency plane ($\mbox{Re}\{s\}<0$). 

\subsection{The spectral Green's function}
\begin{figure}[H]
  \center
  \includegraphics[width=\columnwidth]{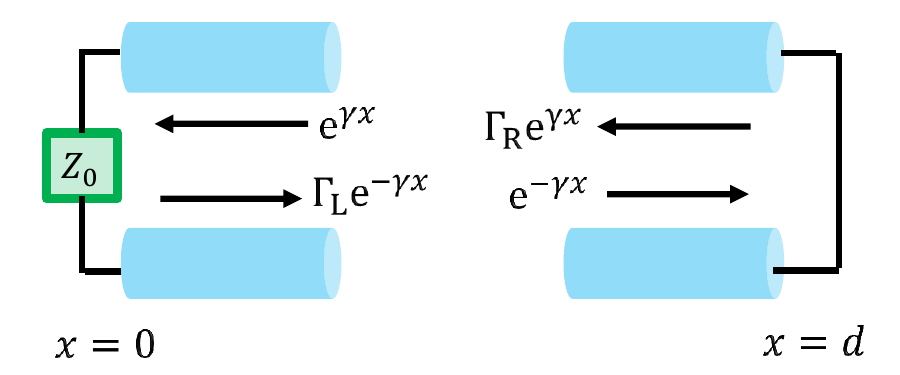}
  \caption{Two half sided infinite transmission line problems terminated on their other hand. By solving the two simple problems Green's function is obtained.}
  \label{Fig.2.}
\end{figure}

The spectral Greens function $g_v(x,x',s)$ is obtained from Eq.\eqref{5} by following a standard procedure \cite{felsen1994radiation},
\begin{eqnarray}\label{9}
  g_v(x,x',s)&=&\frac{F_{v}(x,x',s)}{W(s)},\nonumber\\ F_{v}(x,x',s)&=&\overleftarrow{u}(x^{<})\overrightarrow{u}(x^{>})
\end{eqnarray}
where $\overleftarrow{u}(x)$  [$\overrightarrow{u}(x)$] is the solution of the homogeneous equation corresponding to Eq.~(\ref{5}) that satisfies the boundary conditions at left i.e., at $x=0$ [right, i.e., at $x=d$] (see Fig.~\ref{Fig.2.}), $x^{<}=\min(x,x')$ and $x^{>}=\max(x,x')$. Moreover,  $W(s)$, the Wronskian tune the discontinuity between the left and right solutions, $\overleftarrow{u}(x)$ and  $\overrightarrow{u}(x)$, at the source location $x'$ reads,
\begin{equation}\label{10}
  W \left[ \overleftarrow{u}, \overrightarrow{u} \right]=\overleftarrow{u}(x)\frac{d}{dx}\overrightarrow{u}(x)-\overrightarrow{u}(x)\frac{d}{dx}\overleftarrow{u}(x).
\end{equation}
The homogenous, left and right solutions, i.e.,  $\overleftarrow{u}(x)$ and $\overrightarrow{u}(x)$  are expressed by simple Fresnel coefficients, $\Gamma_{L,R}$,
\begin{eqnarray}\label{11}
  \overleftarrow{u}(x)&=&e^{\gamma x}+\Gamma_L e^{-\gamma x},\nonumber\\
  \overrightarrow{u}(x)&=&e^{-\gamma x}+\Gamma_R e^{\gamma x},
\end{eqnarray}
where $\Gamma_L(s)=[Z_0-Z_2(s)]/[Z_0+Z_2(s)]$ is the reflection coefficient at $x=0$, $Z_2(s)=\sqrt{(sL_2+R_2)/(sC_2+G_2)}$ is the characteristic impedance of the TL model of the layer \emph{after} the temporal switching. Note that the square root should be selected such that $\mbox{Re}\{Z_2\}>0$, and $\Gamma_R=-e^{-2\gamma d}$. Next, by using Eq. \eqref{11} in Eq.~(\ref{10}), the Wronskian is evaluated as follows
\begin{equation}\label{12}
  W(s)=-2\gamma (1+\Gamma_L e^{-2\gamma d}).
\end{equation}
By substituting Eqs.~(\ref{11}) and (\ref{12}) into Eq.\eqref{9} the spectral domain Green function is expressed as a sum of four wave contributions, as follows
\begin{eqnarray}\label{13}
    \label{13b}
    &g_v(x,x',s)=\left[ e^{-\gamma |x-x'|}-e^{\gamma (x+x'-2d)} \right.\nonumber \\
    &\left.+\Gamma_L e^{-\gamma (x+x')}-\Gamma_L e^{\gamma (|x-x'|-2d)} \right]/W(s).
\end{eqnarray}

Inserting Eq.~\eqref{13} into Eq.~\eqref{7} gives the spectral domain voltage $V(x,s)$.

Since the source term in Eq.~\eqref{7}, $f_s(x',s)$, is an analytic function, $g_v(x,x',s)$ and $V(x,s)$ share the same  singular points in the complex $s$ plane.
Then, applying Eq.~\eqref{7} with Eq.~\eqref{13} into Eq.\eqref{8} provides the time domain voltage at any point inside the TL model of the layer,
\begin{equation}\label{New.1}
V(x,t)=\frac{1}{2\pi j} \int_{-j\infty}^{+j\infty}\hspace{-0.2cm} \frac{e^{st}}{W(s)}\int_{-\infty}^{\infty}\hspace{-0.2cm} F_{v}(x,x',s) f_{s,v}(x',s)dx' ds.
\end{equation}

For the completeness of the discussion, a similar procedure is applied to obtain the time domain current $I(x,t)$ (see appendix A).

\section{Going beyond the Rozanov bound}
\label{Sec.Rozanov}
We use the proposed time-variant approach to overcome Rozanov's bound for LTI systems with an ultra-wideband as well as quasi-monochromatic signal, i.e., a modulated short time pulse. Recall that the Rozanov bound is formulated as \cite{rozanov2000ultimate},
\begin{equation}\label{25}
|\ln \rho_{0}| \left(\lambda_{\max}-\lambda_{\min}\right)\leq 2\pi^{2}\mu_{s}d
\end{equation}
where $\rho_{0}$ is the maximal reflection within the operating wavelength band $\left[\lambda_{\min},\lambda_{\max}\right]$ and $\mu_{s}$ denotes the static permeability. Next, we demonstrate by a numerical example how this LTV approach may lead to bypassing Rozanov's bound.

In the following we apply the numerical procedure discussed above for the solution of pulsed wave scattering by a temporally switched layer. We select $t_{s}=0$ as the switching moment where the pulse is located at the center of the layer. The excited voltage and current waves in the TL problem at $t=0^{-}$ were considered in the form of a modulated Blackman type function \cite{nuttall1981some}, and presented in Eq.\eqref{23}. If there exist a phase difference between the baseband and the modulation terms, there will be several modulation frequencies on the ultra-wideband regime where the average of the entire signal is close to zero. This implies that their frequency response is akin to a quasi-monochromatic or narrowband signal. In our example, there is no such phase difference.

\begin{equation}\label{23}
\begin{aligned}
&V_{0}(t)=\frac{1}{w}\sum_{k=0}^{3}a_{k} \cos\left(\frac{2\pi k\left(v_{p}t-\frac{d}{2}\right)}{w}\right) \Pi_{T_{p}}\left(t-\frac{T_{p}}{2}\right)\\
&\hspace{2.7cm}\cos\left(\frac{2\pi f_{0}\left(v_{p}t-\frac{d}{2}\right)}{v_{p}}\right),
\end{aligned}
\end{equation}
where $w$ is the pulse width within the layer, $v_{p}$ is the wavefront propagation speed inside the layer before switching, $\{a_{k}\}$ are Blackman's weighting coefficients, $f_{0}=1/T_0$ is the carrier frequency with $T_0$ being the time period of the carrier signal, and $\Pi_{T_{p}}(t)=H(t+\frac{T_{p}}{2})-H(t-\frac{T_{p}}{2})$, where $H(t)$ is the heaviside function.
We considered the pulse duration $T_{p}=w/v_{p}$. Using these notations, $T_{p}/T_0$ is a measure of the signal's fractional bandwidth. To clarify our notations, note that with $(i)$ $T_{p}/T_0 \to 0$ the signal is practically a baseband signal while with $(ii)$ $T_{p}/T_0  \to \infty$ the signal is practically time-harmonic. In the range in-between, the signal goes from being an ultrawideband for $T_{p}/T_0 \lesssim O(1)$ to quasi-monochromatic and to narrowband for $T_{p}/T_0 \gtrsim O(1)$.

For the demonstration we considered the pulse temporal duration $T_{p}=1.6 [ns]$. The physical layout is that of a layer with $d=0.4[m]$ that initially is characterized by $\epsilon_{1}=1.5\epsilon_{0}$, $\mu_{1}=\mu_{0}$ and $\sigma_{1}=0$, $\sigma_{m1}=0$. At time $t=t_{s}$, while the pulse is tightly contained within the layer (i.e., $w\approx d$) the \emph{permittivity and conductivity are simultaneously switched} to $\epsilon_{2}=0.75\epsilon_{0},\sigma_2>0,\sigma_{m2}=0$. We stress that the layer's permeability remain \emph{unchanged} ($\mu_{2}=\mu_{0}$), and consequently \emph{we pump energy into the wave system} during the switching process.

Next, we evaluate the ratio of the \emph{absorbed energy} in the layer to the incident energy for both time variant and invariant layers, and for different values of $T_{p}/T_0$. To that end, we make use of the reflected energy. The relation between the incident, reflected and absorbed energies are given by
\begin{equation}\label{26}
\frac{E_{\rm ref}}{E_{\rm inc}}=\frac{\int_{-\infty}^{\infty}|\rho_{0}(f) v_{0}(f)|^{2}df}{\int_{-\infty}^{\infty}|v_{0}(f)|^{2}df},\quad \frac{E_{\rm abs}}{E_{\rm inc}}=1-\frac{E_{\rm ref}}{E_{\rm inc}}.
\end{equation}
where $|v_{0}(f)|$ is the magnitude of the fourier transform of the signal ($|v_{0}(f)|=|\mbox{FT}\{V_{0}(t)\}|$),  $E_{\rm inc}$, $E_{\rm ref}$ and $E_{\rm abs}$ denote the incident, reflected and absorbed energies, respectively. It should be noted that generally $E_{\rm ref}$ is composed of two contributions; (1) an early time ($t<t_{s}$) pulsed wave due to impedance mismatching at the layer's boundary and (2) a back reflected wave due to the switching of the TL parameters and the multiple wave bounces in the absorbing layer for $t>t_{s}$.

First, in LTI case where the layer is time invariant and Rozanov's bound exists, the parameter $\rho_{0}$ in Eq.\eqref{25} is taken to maximize the left hand side of Eq.~\eqref{25} within the operating band to give
\begin{equation}
\label{27}
\rho_{0} = \begin{cases} e^{\frac{-2\pi^{2}\mu_{s}d}{\Delta \lambda}} &\mbox{if } \lambda \in  \left[\lambda_{\min},\lambda_{\max}\right] \\
1 & \mbox{else }  \end{cases}
\end{equation}
where $\Delta \lambda = \lambda_{\max}-\lambda_{\min}$. Note that $\rho_{0}$ is a function of $\Delta \lambda$ which can be expressed by two frequency related parameters: a center frequency $f_{c}$ and the bandwidth $BW$ as follows: $\Delta \lambda={c_{0} BW}/\left[{(f_{c}-{BW}/{2})(f_{c}+{BW}/{2})}\right]$ and $c_{0}$ is the speed of light in vacuum. In order to minimize the reflection within the operating band, we preform a sweep of the parameters $f_{c}$ and $BW$ to maximize the ratio $E_{\rm abs}/E_{\rm inc}$ in Eq.~\eqref{26}. If the modulation frequency of the incident signal is much larger than its bandwidth, i.e. a narrowband signal, the optimal frequency will be equal to the modulation frequency. However, for quasi-monochromatic and ultra-wideband signals this is not always the case, therefore a sweep along $f_{c}$ and $BW$ is preformed.

In contrast, the absorbed energy of the \emph{LTV system} is evaluated by subtracting the total reflected energy from the incident energy, i.e. $E_{\rm abs}=E_{\rm inc}-E_{\rm ref,initial}-E_{\rm ref,s}$, where $E_{\rm ref,initial}$ is the initial reflected energy due to impedance mismatch before switching ($t_{0}\leq t \leq t_{0}+T_{p} $) and $E_{\rm ref,s}$ denotes the total reflected energy after switching ($t>t_{s}$).

\begin{figure}[H]
     \centering
    \begin{subfigure}[b]{0.48\columnwidth}
         \centering
         \includegraphics[width=\textwidth]{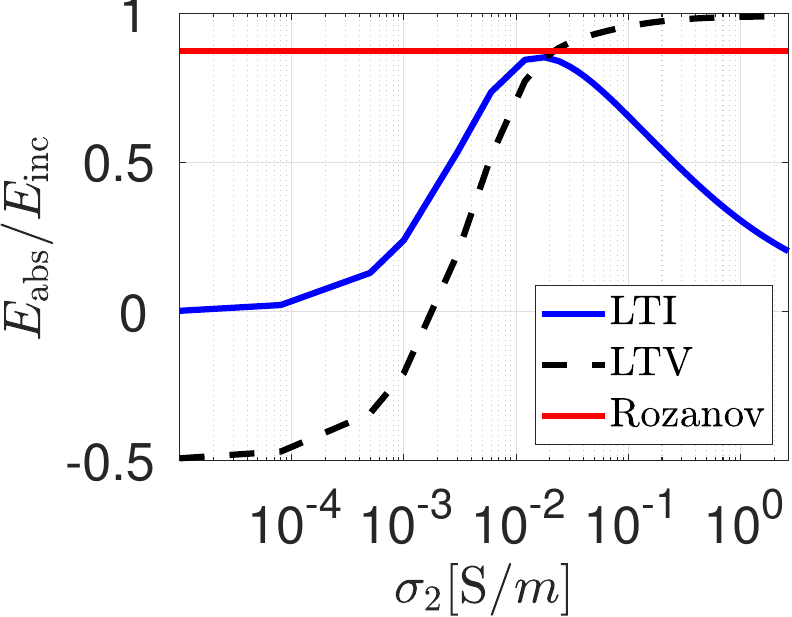}
         \caption{$T_{p}/T_0\rightarrow0$}
         \label{woT_0_0}
    \end{subfigure}
    \hfill
    \begin{subfigure}[b]{0.48\columnwidth}
         \centering
         \includegraphics[width=\textwidth]{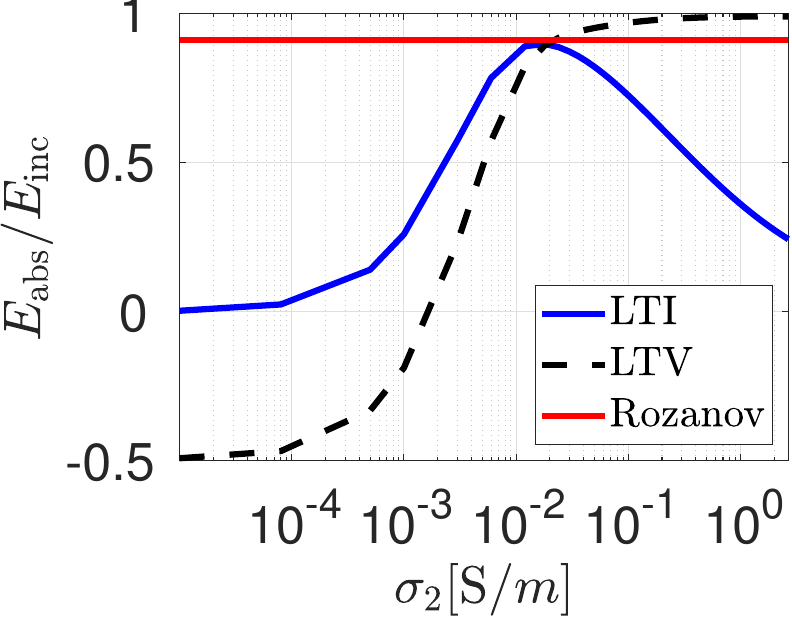}
         \caption{$T_{p}/T_0=0.992$}
         \label{woT_0_992}
    \end{subfigure}
\\
    \begin{subfigure}[b]{0.48\columnwidth}
         \centering
         \includegraphics[width=\textwidth]{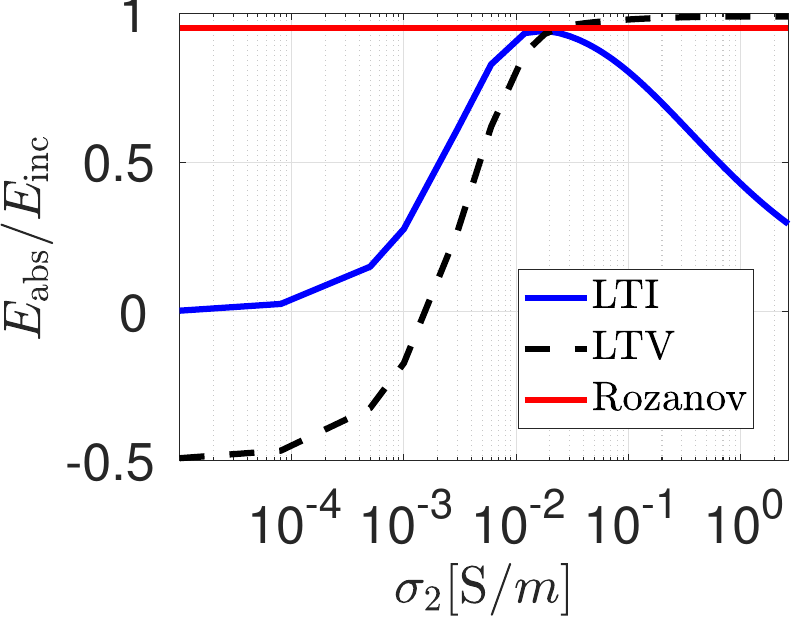}
         \caption{$T_{p}/T_0=1.472$}
         \label{woT_0_1472}
    \end{subfigure}     \hfill
    \begin{subfigure}[b]{0.48\columnwidth}
         \centering
         \includegraphics[width=\textwidth]{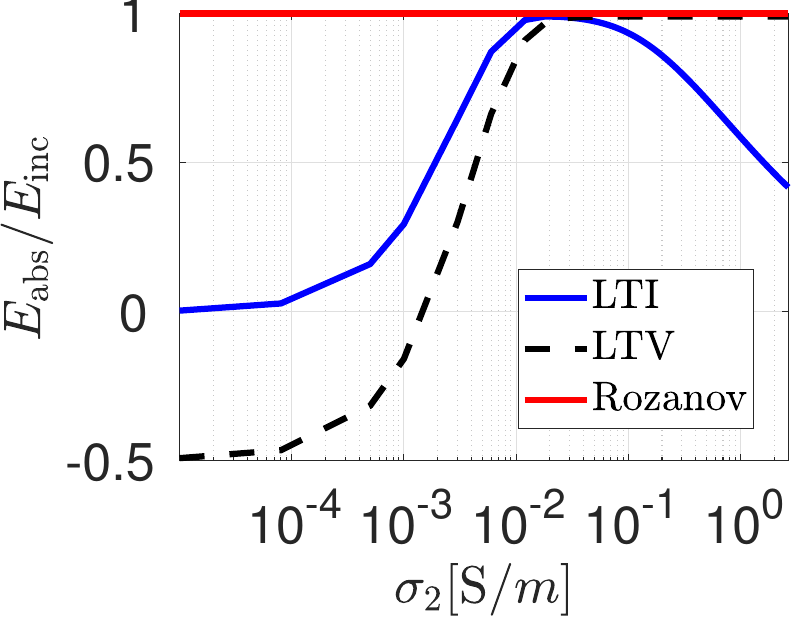}
         \caption{$T_{p}/T_0=2.88$}
         \label{woT_0_2880}
    \end{subfigure}
\caption{Absorption caused by time invariant and time variant system compared to Rozanov's bound as a function of the conductivity and for some values of the signal's frequency bandwidth.}
\label{Fig.7.}
\end{figure}

\begin{figure}[H]
     \centering
     \begin{subfigure}[b]{0.48\columnwidth}
         \centering
         \includegraphics[width=\textwidth]{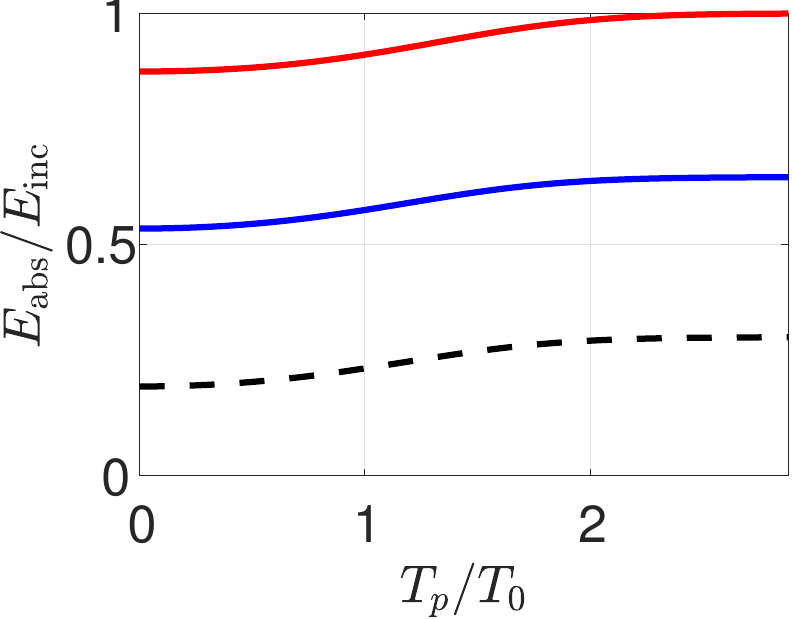}
         \caption{$\sigma=0.003 [\rm S/m]$}
         \label{sgm_3}
     \end{subfigure}
     \hfill
     \begin{subfigure}[b]{0.48\columnwidth}
         \centering
         \includegraphics[width=\textwidth]{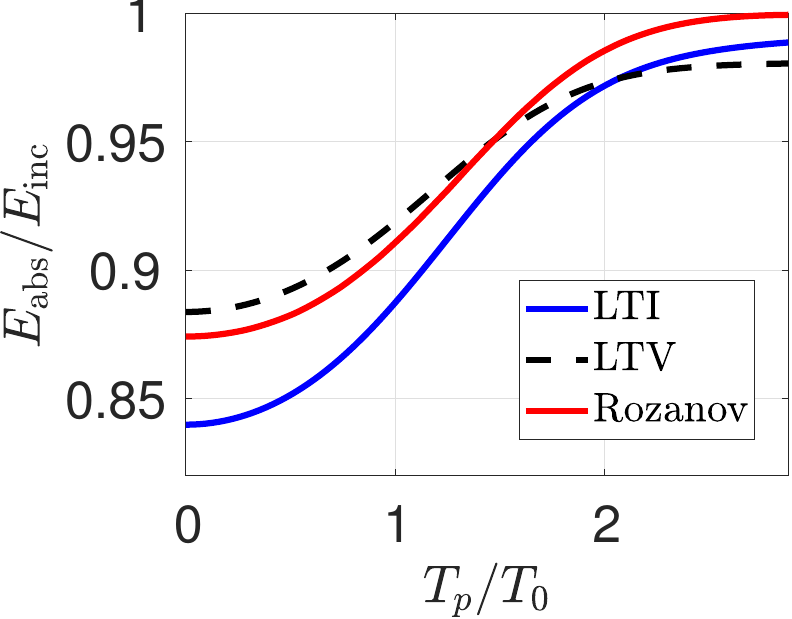}
         \caption{$\sigma=0.024 [\rm S/m]$}
         \label{sgm_24}
     \end{subfigure}
\\
     \begin{subfigure}[b]{0.48\columnwidth}
         \centering
         \includegraphics[width=\textwidth]{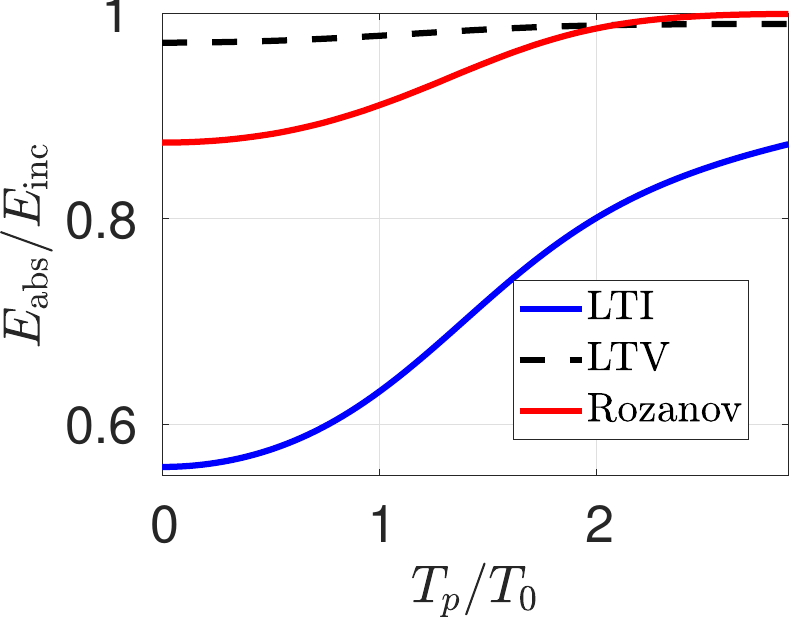}
         \caption{$\sigma=0.18 [\rm S/m]$}
         \label{sgm_180}
     \end{subfigure}
     \hfill
     \begin{subfigure}[b]{0.48\columnwidth}
         \centering
         \includegraphics[width=\textwidth]{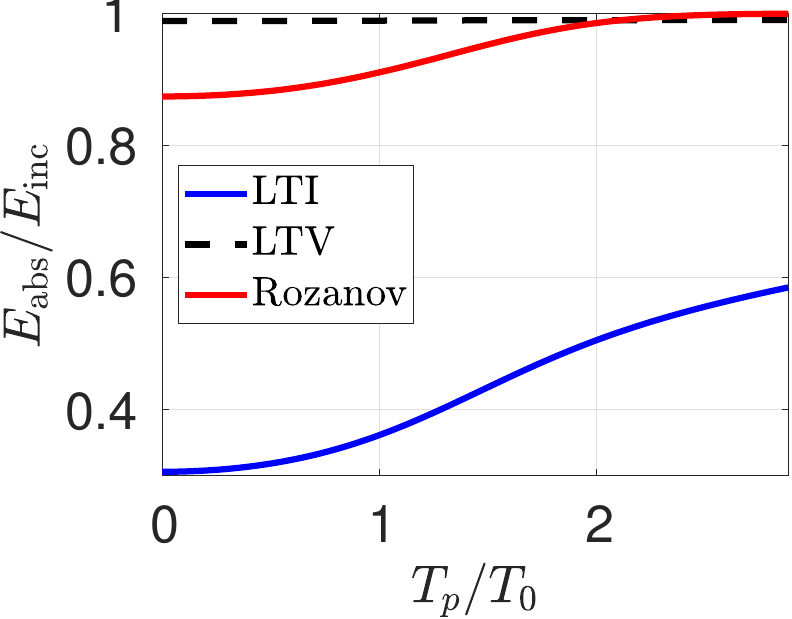}
         \caption{$\sigma=1.002 [\rm S/m]$}
         \label{sgm_1002}
     \end{subfigure}

        \caption{Absorption caused by time invariant and time variant schemes compared to Rozanov's bound as a function of the signal's bandwidth and for some values of the conductivity.}
        \label{Fig.9.}
\end{figure}

\begin{figure}[H]
  \center
  \includegraphics[width=1\columnwidth]{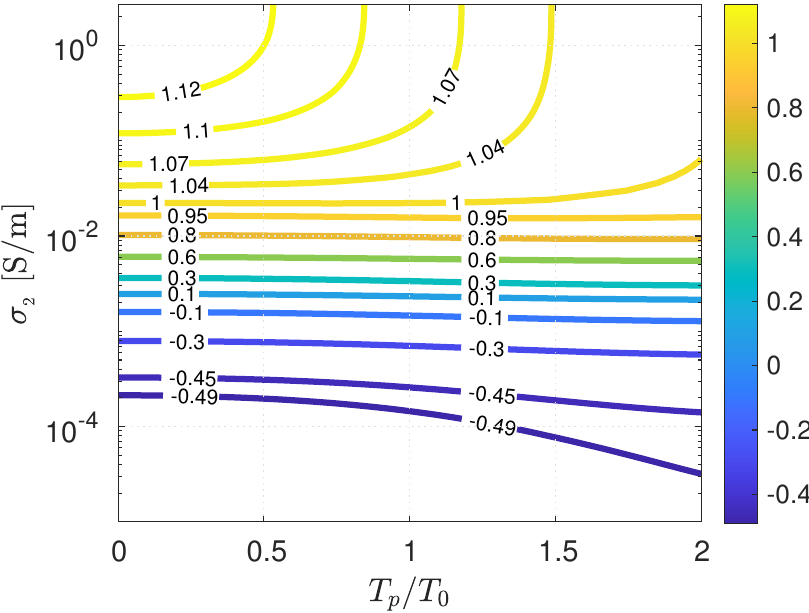}
  \caption{A contour plot (``isolines'') of the ratio of the absorbed energy of the LTV scheme to that predicted by Rozanov's bound, i.e., $(E_{\rm{abs}})\big |_{\rm{LTV}}/(E_{\rm{abs}}) \big |_{\rm{Rozanov}}$, (see in Eq.~\eqref{26}).}
  \label{Fig.10.}
\end{figure}

The absorption optimization results are presented in Fig.~\ref{Fig.7.} as a function of the conductivity $\sigma_2$, for several values of $T_{p}/T_0\in(0,3)$ covering the range of extreme ultra-wideband to quasi-monchromatic signals. In Fig.~\ref{Fig.7.}, the red lines represent the proper Rozanov's bound, the blue lines represent a LTI system absorption response, and the black dashed lines are used to depict the absorption with the time-varying scheme. It can be observed in Fig.~\ref{Fig.7.}  that for any given value of $T_{p}/T_0$ (signal bandwidth) it is possible to determine a range of conductivities $\sigma_2$ such that the time-varying scheme exhibits better, or at least equal, performance in comparison to Rozanov's bound. In particular, for certain bandwidth regimes, an enhanced performances of the time-varying scheme occur for $\sigma \gtrsim 0.024$  (Figs.~\ref{woT_0_0},~\ref{woT_0_992}~\ref{woT_0_1472}) while for other regimes it asymptotically occurs for relatively larger values of the conductivity. Obviously, for the LTI case, shown in blue lines,  the performance is always below the bound.\\
A complementary view of the enhanced performances of the LTV scheme over Rozanov's bound is demonstrated in Fig.~\ref{Fig.9.} that depicts the LTI, LTV and Rozanov's bounds in (blue, dashed black and red lines, respectively) for $\sigma_2=0.03, 0.054, 0.18, 1.002 [\rm S/m]$ as a function of the relative bandwidth $T_{p}/T_0$. It can be easily noted by Fig.~\ref{Fig.9.} that for $\sigma_2 \gtrsim 0.024[\rm S/m]$ the LTV scheme becomes better than Rozanov's bounds in some bandwidth regimes of the excitation signal (for the parameters used here, $\sigma_2\approx0.024[\rm S/m]$   can be identified as the minimal required value of loss for the LTV scheme in order to bypass the Rozanov's bound for ultrawideband / quasi-monochromatic signal excitation). This enhanced performance in terms of better absorption and extended bandwidth improves as $\sigma_2$ is moderately increased.
Finally, Fig.~\ref{Fig.10.} depicts a contour plot (``isolines'') of the ratio between the absorbed energy of the LTV scheme and that predicted by the Rozanov's bound, i.e., \[\frac{(E_{\rm{abs}})\big |_{\rm{LTV}}}{(E_{\rm{abs}}) \big |_{\rm{Rozanov}}}.\]Note that the results presented in Fig.~\ref{Fig.7.} and Fig.~\ref{Fig.9.} are obtained by horizontal and vertical ``cuts'' in Fig.~\ref{Fig.10.} along cartesian grid lines. Furthermore the extent of the frequency bandwidth ``windows'' for which the LTV scheme bypass the Rozanov bound are easily spotted by observing the regions in which the contour line hight is larger than unity, i.e., ($(E_{\rm{abs}})\big |_{\rm{LTV}}/(E_{\rm{abs}}) \big |_{\rm{Rozanov}}>1$), for $\sigma_{2} \gtrsim 0.024$ and $T_{p}/T_0<1.9$ while for $T_{p}/T_0>1.9$ it asymptotically approaches 1. The negative values in Fig.~\ref{Fig.10.} for low values of conductivity ($\sigma_{2}$), are due to the energy injection into the wave system during the switching. Therefore, the reflected field is stronger than the incident field and can be recognized as negative absorption.
To conclude the discussion, this example demonstrates that Rozanov's absorption bound is valid only for LTI systems, and can be bypassed for certain cases by utilizing mechanisms due to time variation, moreover the example demonstrates the appealing characteristics of time-varying scheme that involves \emph{permittivity and conductivity (material loss) switching},  for ultra-wideband and quasi-monochromatic as well as narrowband (time harmonic) signals.
In order to reduce  the ``break-even'' points for which the Rozanov's bound bypasses the LTV scheme in Fig.~\ref{Fig.9.}, and to further lower  the required minimal $\sigma_2$ values, it may be beneficial to use the more complex time-variation schemes that  involve, besides permittivity and conductivity switching,  also the switching of the material  permeability.
The bounds demonstrated in this discussion were evaluated using the absorbed energy spectrum of the signal along the \emph{entire frequency band} (note that $f \in (-\infty,\infty)$ in Eq.~(\ref{26})). It should be recalled that switching the medium permeability and permittivity, changes the pulse temporal width and consequently its frequency spectrum \cite{shlivinski2018beyond}. Therefore, if the interest is only, for example, in the minimization of the power reflection in the frequency band of the incident wave, while discarding the reflection outside this band, then the time-switching of the  permittivity  only can be applied such that parts of the frequency spectrum are disperse to other bands outside that of the incident wave. Such an effect is utilized and nicely discussed in  \cite{li2020temporal} for quasi-monochromatic signals.

\section{Realistic Implementation}
In the previous section we have demonstrated a theoretical venue to bypass Rozanov's bound by performing an abrupt switching of both conductivity and permittivity of the dielectric layer. There are however two important, challenges with this scheme. First, we have to design a metamaterial and as simple as possible switching scheme that will change the effective properties (conductivity and permittivity), between the values that were suggested in section IV. Second, abrupt switching is impractical, and even may contradict causality, and therefore it is essential to demonstrate that the proposed scheme to bypass Rozanov's bound applies also with gradual switching. These issues are addressed bellow.

\subsection{Effective Media}
\begin{figure}[H]
  \center
  \includegraphics[scale=0.6]{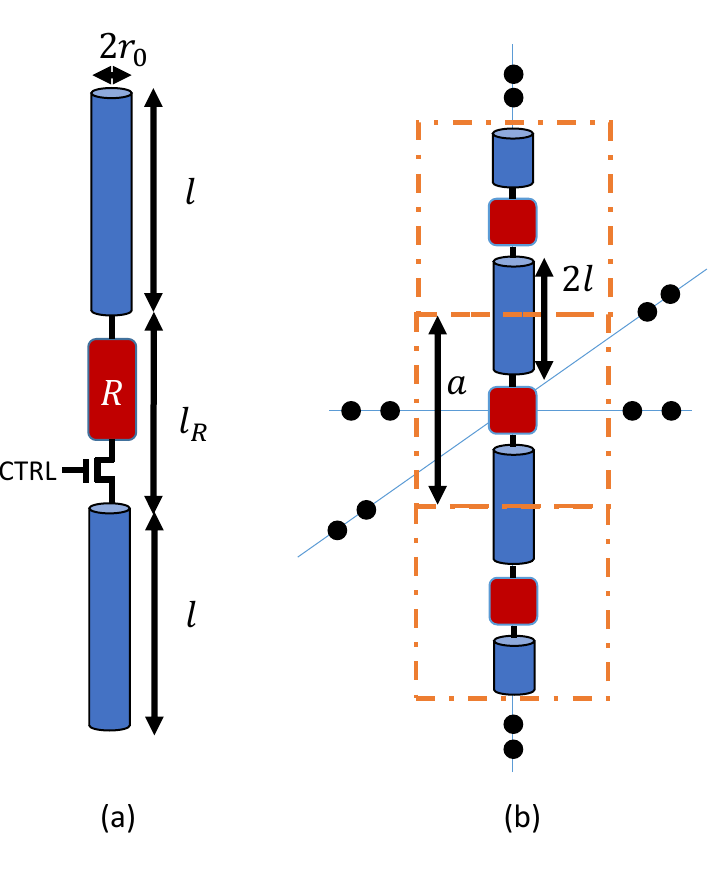}
  \caption{Metamaterial topology. (a) Each element is composed of a dipole connected by a resistor in its two terminals. The dipole has a total length of $2l$ and a thickness of $2r_{0}$. The length of the resistor $R$ is $l_{R}$. The switching device is composed of a single CMOS transistor with a controlling feature(CTRL). (b) Each unit cell (sized a) is attached to its neighbors vertically. By changing the resistance to a finite value, the current flows vertically across the unit cells, thus altering the 3D system to a 2D system. }
  \label{Fig.unit cell.}
\end{figure}

Here, we suggest a practical system that realizes the transition \{$\epsilon_{1}=1.5\epsilon_{0}, \sigma_{1}=0$\} to \{$\epsilon_{2}=0.75\epsilon_{0}, \sigma_{2}=0.18 [S/m]$\} demonstrated in Fig.~\ref{sgm_180}. Such extreme transition between a pure dielectric material to an absorbing material with high losses requires a variation of the topology of the metamaterial during the switching. It can be achieved by altering a 3D system to a 2D system. As previously shown, this transition achieves $(E_{\rm{abs}})\big |_{\rm{LTV}}/(E_{\rm{abs}}) \big |_{\rm{Rozanov}}>1$, for an increased range of modulated pulses. We use a metamaterial structure that is composed of cubical unit cells with unit cell size $a$. Each cell contains a PEC wire (dipole) with length of each arm denoted by $l$ and $l_{R}$ denotes the resistor length. The total dipole length is $2l+l_{R}$, and this should be equal to $a$ since $a$ is the cubical unit cell size. The dipole's diameter is denoted by $2r_{0}$ loaded with a resistor $R$ between its two terminals (see Fig.~\ref{Fig.unit cell.}). We obtain the required effective properties ($\epsilon_{1},\sigma_{1}$),($\epsilon_{2},\sigma_{2}$) by proper selection of the dimensions of the wires and the resistance \cite{serdyukov2001electromagnetics,tretyakov2003analytical,tretyakov1995maxwell}. We considered the following parameters: $a=20 [mm],r_{0}=0.846[mm]$,$l_{R}=7.4[m]$. The wires are located in a host dielectric media with $\epsilon_{h}=1.3$ (such as CarbonFoam PU 2 R PU \cite{Material}).

\begin{figure}[H]
  \center
  \includegraphics[width=1\columnwidth]{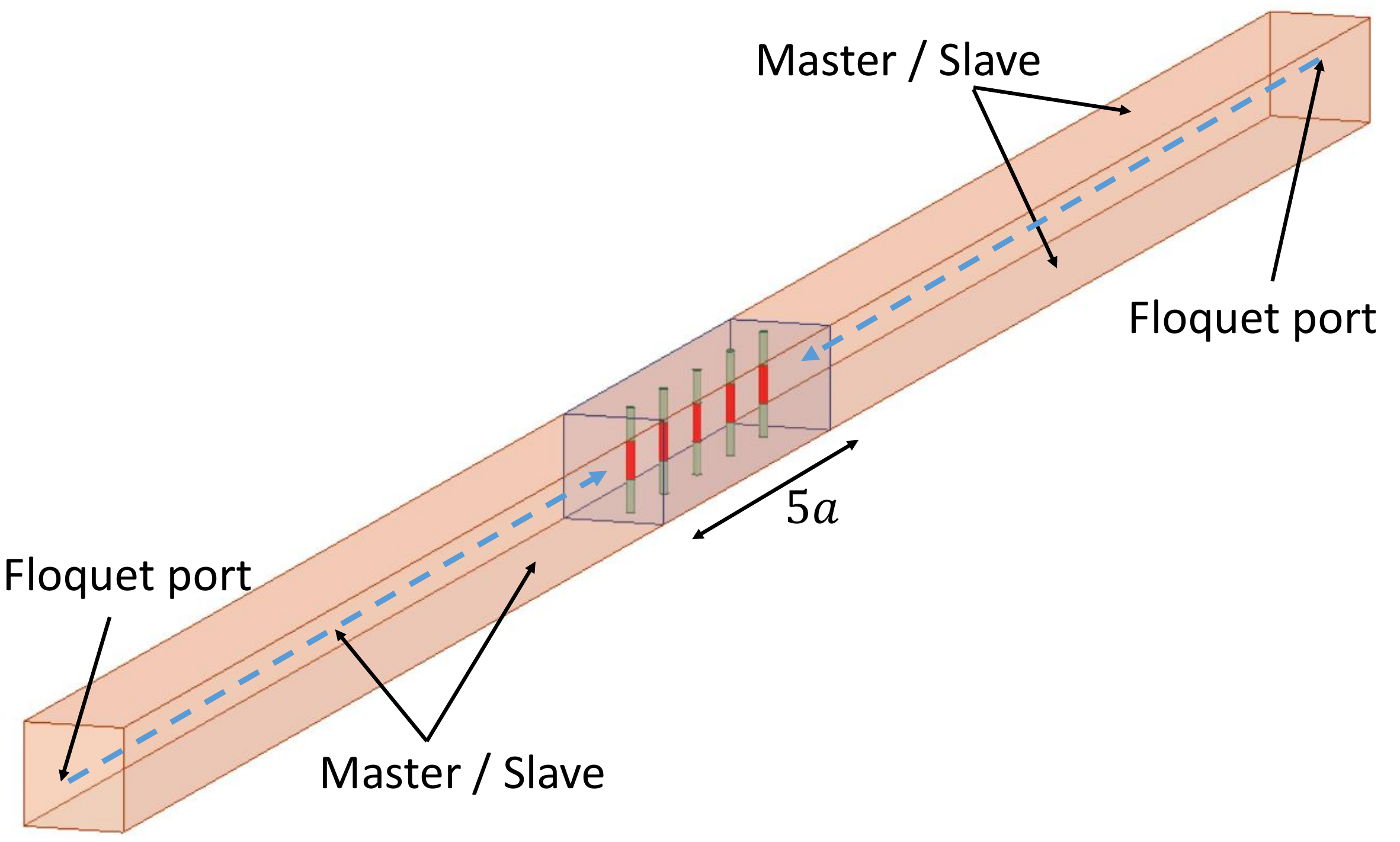}
  \caption{HFSS simulation setup. The periodic structure (Master Slave boundary conditions) is excited by a Floquet port where only the fundamental mode is propagating. By evaluating the $\bar{\bar{S}}$ parameters and by performing a post-processing de-embedding we extract the effective properties of the discrete slab. }
  \label{Fig_Sim_Setup}
\end{figure}

\begin{figure}[H]
     \centering
     \begin{subfigure}[b]{0.48\columnwidth}
         \centering
         \includegraphics[width=\textwidth]{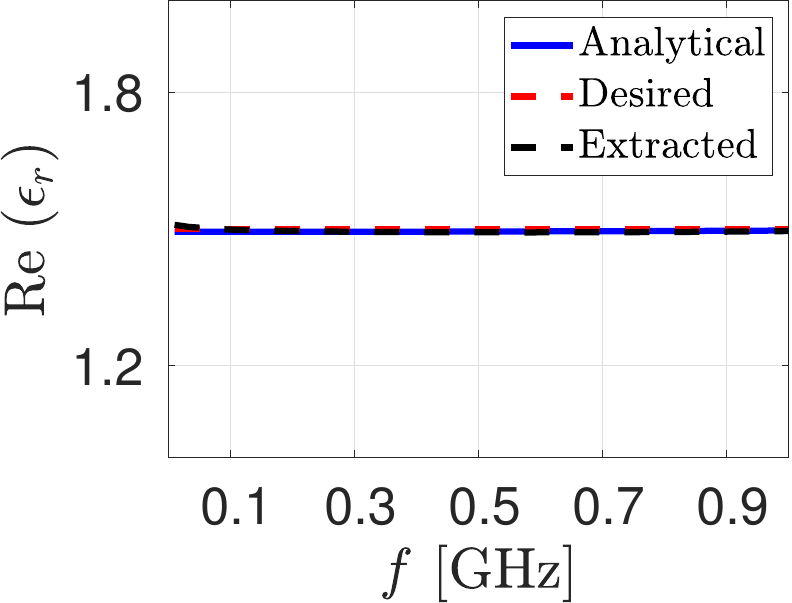}
         \caption{}
         \label{Eps_Re_Before}
     \end{subfigure}
     \hfill
     \begin{subfigure}[b]{0.48\columnwidth}
         \centering
         \includegraphics[width=\textwidth]{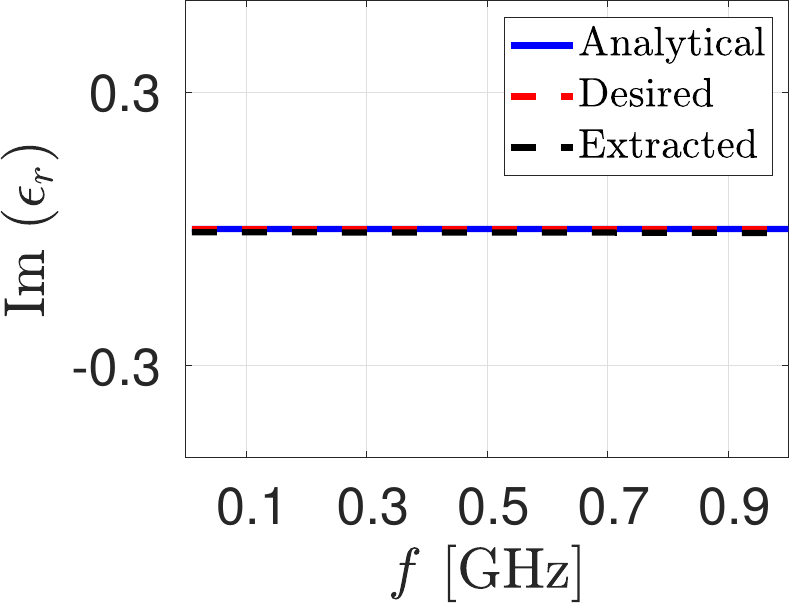}
         \caption{}
         \label{Eps_Im_Before}
     \end{subfigure}
\\
     \begin{subfigure}[b]{0.48\columnwidth}
         \centering
         \includegraphics[width=\textwidth]{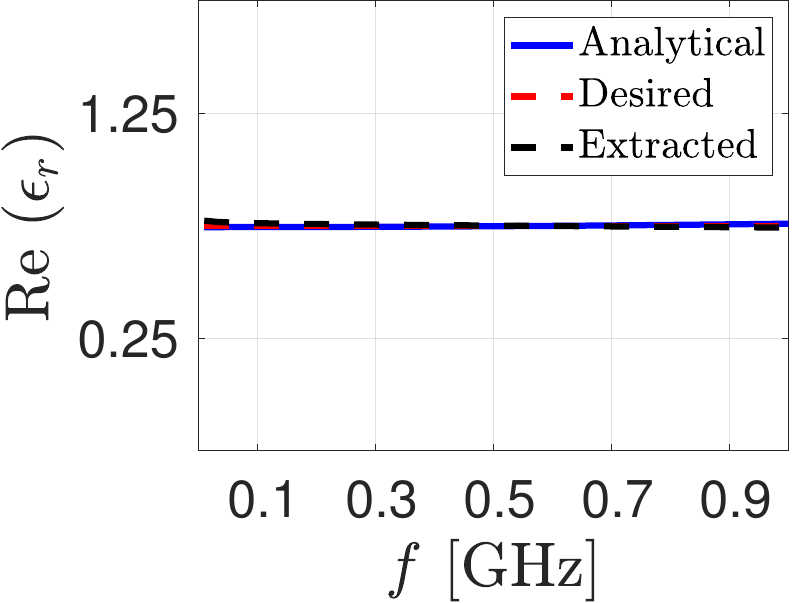}
         \caption{}
         \label{Eps_Re_After}
     \end{subfigure}
     \hfill
     \begin{subfigure}[b]{0.48\columnwidth}
         \centering
         \includegraphics[width=\textwidth]{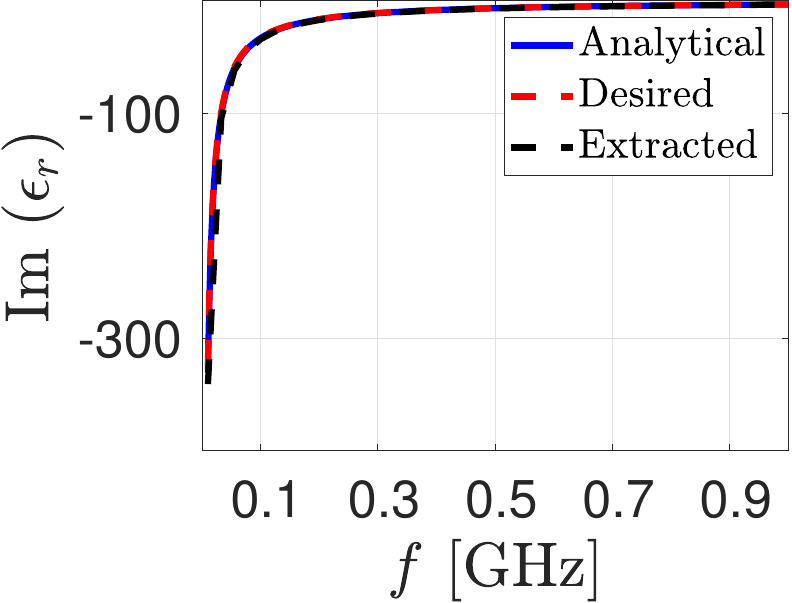}
         \caption{}
         \label{Eps_Im_After}
     \end{subfigure}

        \caption{Analytical, desired and extracted  complex $\epsilon$ as a function of frequency $f$. (a)-(b) before switching, (c)-(d) after switching.}
        \label{Fig_Eps_Comp}
\end{figure}

\emph{Effective media characteristics before switching} are $\epsilon_{1}=1.5\epsilon_{0},\sigma_{1}=0$.  Here, an open circuit ($R\rightarrow \infty$) is enforced between the two terminals, so the electric polarizability of a single wire is given by $\alpha_{ee}=l^{2}/\left(j2\pi fZ_{\rm {wire}}\right)$, where $Z_{\rm {wire}}$ is the input impedance of the wire \cite{serdyukov2001electromagnetics,king1969antennas},

\begin{equation}
\label{Z_wire}
Z_{\rm {wire}}=\frac{\eta_{0}\Psi_{\rm{dr}}}{2\pi j\left[1+k^{2}l^{2}F/3-jk^{3}l^{3}\frac{1}{3(\Omega-3)}\right]},
\end{equation}
where the dimensionless parameters are $F=1+1.08/(\Omega-3)$, $\Omega=2\ln\left(2l/r_{0}\right)$ and $\Psi_{\rm{dr}}=2\ln\left(l/r_{0}\right)-2$.
The dipole polarizability in free space should satisfy the radiation condition, therefore a radiation correction should be taken into account. However, for non-resonant, quasi-static metamaterial such correction will produce a minor impact on the effective properties, since $|\mbox{Re}(\alpha)|>>|\mbox{Im}(\alpha)|$ (see Appendix B for further discussion). Applying the electric polarizability $\alpha_{ee}$ with the Maxwell Garnett homogenization procedure \cite{tretyakov1995maxwell} for 3D structures provides the analytic complex effective permittivity,

\begin{equation}
\label{28}
\epsilon_{\rm{eff}}=\epsilon_{0}\epsilon_{h}+N\alpha_{ee}\left(1-\frac{N\alpha_{ee}}{3\epsilon_{0}\epsilon_{h}}\right)^{-1},
\end{equation}

where $N=1/a^{3}$ denotes the lattice density. Using this analytical formalism with $l=0.0085[m]$ provides $\epsilon_{\rm{eff}}\sim\epsilon_{1}$.
In order to verify and fine tune the analytic results, we used HFSS to extract the effective parameters of the homogenized slab. We used a well known extraction technique \cite{smith2002determination} which uses the scattering ($\bar{\bar{S}}$) parameters of a normal incidence illumination (see Fig.~\ref{Fig_Sim_Setup} for the simulation setup). A post-processing de-embedding was performed to obtain the $\bar{\bar{S}}$ parameters at the interface between the slab and the vacuum (marked by blue dashed arrows). By the simulation optimization, the following design parameters were obtained in order to fit with the desired effective material properties before the switching: $l=0.0063[m],R=300\times10^{9} \left[\Omega\right]$. Due to edge effects, there is a slightly deviation between the optimal dipole length in compare to the analytical formulas which is expectable under real design considerations.
Results are presented in Figs.~\ref{Eps_Re_Before},~\ref{Eps_Im_Before} with the standard definition of the relative permittivity, $\epsilon_{r}=\epsilon_{\rm{eff}}/\epsilon_{0}$, and $\epsilon_{r}=\epsilon_{r}^{'}-j\sigma/\left(2\pi f \epsilon_{0}\right)$, where $\epsilon_{r}^{'}$ denotes the real part of $\epsilon_{r}$.

\emph{Effective media characteristics after switching} are $\epsilon_{2}=0.75\epsilon_{0},\sigma_{2}=0.18[S/m]$.  Here, we use a 2D homogenization procedure suggested in \cite{tretyakov2003analytical} which was developed for loaded wires structure,
\begin{equation}
\label{29}
\epsilon_{\rm{eff}}=\epsilon_{0}\epsilon_{h}-\frac{1}{\omega^{2}a^{2}\tilde{L}-j\omega a^{2}\tilde{R}},
\end{equation}
 where $\tilde{L}=\mu_{0}/(2\pi) \ln \left[a^{2}/(4r_{0}(a-r_{0}))\right]$ and $\tilde{R}=R/a \ [\Omega/m]$ are the inductance and resistance per unit length, respectively. In this case, the optimal performances to obtain an homogenized medium with $\epsilon_{2}=0.75\epsilon_{0}, \sigma_{2}=0.18[S/m]$ are obtained with $R=272[\Omega]$ in the theoretical calculations, and $R=264[\Omega]$ extracted by the numerical HFSS optimization. The theoretical calculations assume a lumped resistor, while the simulation also takes into account the dimensions of the resistor. Results are presented in Figs.~\ref{Eps_Re_After},~\ref{Eps_Im_After}. During switching between the two states only the resistance is altered, so by using a switch as shown in Fig. 7a, this switch, during the transition from open to close alters the nature of the metastructure system from 3D to 2D. This change enables to obtain lower values of $\epsilon$ ($\epsilon_{2}<\epsilon_{1}$), while considering an additional loss mechanism($\sigma_{2}$).
An excellent agreement between desired, analytical and extracted effective parameters for both before and after switching is provided by this scheme as can be observed in Fig.~\ref{Fig_Eps_Comp}. To conclude the discussion, both analytical and numerical calculations show that a dipole based metamaterial can exhibit the required effective media properties for the design of an absorbing material with better performance than those of Rozanov's bound.\\\\

\subsection{Gradual Switching}
\begin{figure}[H]
  \center
  \includegraphics[width=1\columnwidth]{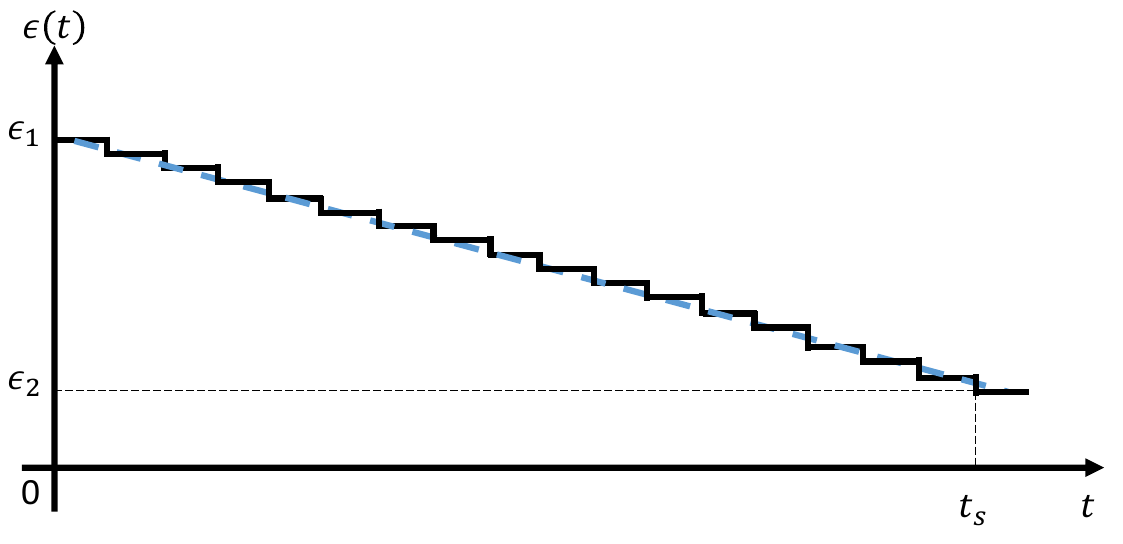}
  \caption{Illustration of the gradual (soft) switching. The switching process begins at $t=0$ and ends at $t_{s}$. It is implemented as a series of ``small sized'' abrupt switchings, with constant steps duration (black). The continuous limit is shown in dashed blue.}
  \label{steps}
\end{figure}

In section IV we demonstrated how the Rozanov bound can be bypassed by using an abrupt switching of the effective parameters. Such a switching scheme may be challenging practically and may even contradict causality. Therefore we consider here a gradual (soft) switching scenario that last over an extended period of time, even larger than the temporal width of the impinging pulse. The switching process begins at $t=0$ while the pulse is contained  within the layer and ends at some time $t_{s}>0$ (see Fig.~\ref{steps}).
In a previous work \cite{hadad2020soft} using WKB analysis some of the authors of this paper have showed the equivalence of soft switching and the staircase approximation. Hence, here, the gradual switching is implemented as a series of ``small sized'' abrupt switchings (staircase steps), where the values of the permittivity and the conductivity change.
In each time step, we evaluate numerically the voltage and current along the TL using the initial conditions that were obtained in the previous step and by considering the effect of the abrupt voltage change due to the switching (see section III). Generally, at each of the small switching steps we require the continuity of the magnetic flux and the electric charge on the TL section that undergoes switching at $x>0^+$, and consequently, the continuity of the voltage and current there is obviously not satisfied. We stress, however, that at $x=0^+$, at the connection point between the TL sections in $x>0$ and $x<0$, this continuity condition does not hold. Instead, at $x=0^+$,  we enforce the continuity of the voltage and the current. This is essential because the properties of the TL section at $x<0$ are unaltered during the entire switching process, implying that also the current and voltage on it remain continuous at each one of the switching steps. Thus at $x=0$ at any switching moment we have,
\begin{eqnarray}\label{Gradual_0}
I_{0}'(0)=\lim_{h \downarrow 0}  \frac{I_{0}(h)-V_{0}(0)/Z_{0}}{h}\nonumber\\
V_{0}'(0)=\lim_{h \downarrow 0}  \frac{V_{0}(h)-Z_{0}I_{0}(0)}{h}.
\end{eqnarray}
Note that in the example, the current remains continuous since the magnetic properties are unchanged. We emphasize that for the gradual switching process (unlike a single abrupt switching), one must evaluate both the voltage and the current in each step, since they are used to evaluate the initial conditions for the next switching ($I_{0}$ and $V_{0}$).

\begin{figure}[H]
  \center
  \includegraphics[width=1\columnwidth]{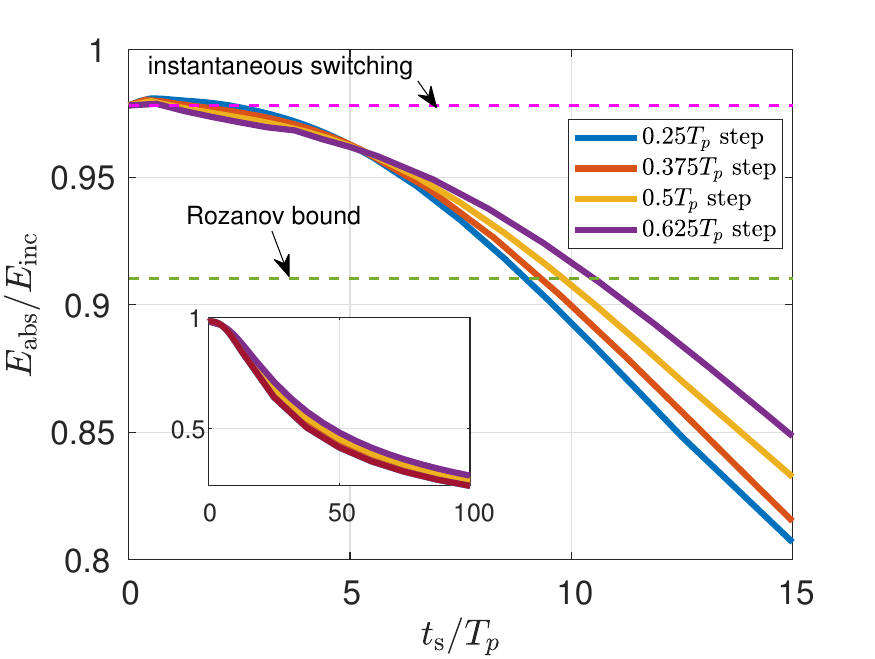}
  \caption{Switching is performed gradually in a staircase manner. In each step, the permittivity and conductivity are abruptly switched. Time steps: blue line - $0.25T_{p}$, red line - $0.375T_{p}$, yellow line - $0.5T_{p}$ and purple line - $0.625T_{p}$. Remarkably, the time-varying absorber under gradual switching performs similarly to what is expected by instantaneous switching, and even better under certain conditions. As the switching time extends above $\sim10T_p$, eventually, the time-invariant does not introduce an improvement compare with the Rozanov bound for LTI absorbers}
  \label{Gradual}
\end{figure}

Here, we consider the transition \{$\epsilon_{1}=1.5\epsilon_{0}, \sigma_{1}=0$\} to \{$\epsilon_{2}=0.75\epsilon_{0}, \sigma_{2}=0.18 [S/m]$\} for $T_{p}/T_{0}=0.992$ (compare to Fig.~\ref{woT_0_992} for a single abrupt switching), with the parameters obtained in the metamaterial realization that was discussed in Sec.V.1.
The absorbtion results are presented in Fig.~\ref{Gradual} for several values of time steps (blue line - $0.25T_{p}$, red line - $0.375T_{p}$, yellow line - $0.5T_{p}$ and purple line - $0.625T_{p}$).
It can be observed by  Fig.~\ref{Gradual} that for switching times that are in the order, and even larger, than the pulse temporal width $t_s<10T_p$, the time variant system provides better performance in compare to Rozanov's bound. Obviously, as the switching time tends to infinity, the system is not time-varying anymore, it remains at its first state where the slab is lossless with $\epsilon_1=1.5\epsilon_0$, and therefore its absorption is zero, as expected, and shown in Fig.~\ref{Gradual}. Remarkably, even switching of 5 times pulse width leads to minor effect on the performance, and interestingly, gradual switching may lead to better performance even when comparing to the case of abrupt switching. This is a consequence of the reduced reflection coefficient caused by a gradual time switching, in comparison to that caused by an instantaneous switching \cite{hadad2020soft}. Consequently, in light of the possibility to use such extended switching times that are substantally larger than the pulse temporal width, the pulse, in these cases, may be either fully or only partially contained within the layer at $t=0$, as the switching process starts, and with no substantial degredation in performance. Thus, increasing the tolerance in the switching process.

\section{Summary and Conclusion}
In this manuscript, we consider the problem of a short pulse propagating and scattering from an electromagnetic wave absorber with time varying properties. A transmission line equivalent was used to formulate an initial boundary problem. The problem was solved by finding the spectral (on the Laplace $s$-plane) Green's function, followed by an inverse Laplace transform. This formalism was used to obtain the time domain reflected voltage and current from which all other required quantities are calculated.
Using this numerical procedure we have solved the problem of reflection by a dielectric layer absorber in which the permittivity and conductance are switched in time. We further demonstrated that Rozanov's LTI absorbtion bound, when calculated over the \emph{whole signal's spectrum range}, can be bypassed for modulated pulsed signals,  ultra-wideband, quasi-monochromatic, and narrowband by such abrupt switching. We presented a time dependent switchable metamaterial structure which exhibits the required effective media properties before and after the switching. Furthermore, we examined a gradual switching that also results in an improved performance in compare to Rozanov's bound, similar, and even better than what is expected by using an idealistic instantaneous switching.

\section*{Acknowledgment}

C. F.  would like to thank to the Darom Scholarships and High-tech, Bio-tech and Chemo-tech Scholarships and to Yaakov ben-Yitzhak Hacohen excellence Scholarship. This research was supported by the Israel Science Foundation (grant No. 1353/19).

\appendix
\section{Time domain current}

Spectral domain Helmholtz equation for the current in a lossy medium can be written as
\begin{subequations}\label{a.1}
  \begin{eqnarray}
  \label{a.1.1}
  &&\frac{\partial^2 I(x,s)}{\partial x^2}-\gamma ^{2}I(x,s)=f_{s,i}(x,s)\\[1ex]
  \label{a.1.2}
  &&f_{s,i}(x,s)=C\frac{\partial V_0(x)}{\partial x}-L(sC+G)I_0(x),
  \end{eqnarray}
\end{subequations}
where $\gamma$ is the complex propagation term and $L,C,R,G$ are the TL inductance, capacitance, series resistance and shunt admittance per unit length, respectively.
Green's function for the current term is obtained in similar way to the voltage (see section III),

\begin{subequations}\label{a.2}
\begin{eqnarray}
  &&g_i(x,x',s)=\frac{F_{i}(x,x',s)}{W(s)}\\
  &&F_{i}(x,x',s)=e^{-\gamma |x-x'|}+e^{\gamma (x+x'-2d)}\\
  &&\hspace{2cm}-\Gamma_L e^{-\gamma (x+x')} -\Gamma_L e^{\gamma (|x-x'|-2d)} \nonumber
\end{eqnarray}
\end{subequations}
where $W(s)$ is the wronskian term (see Eq.\eqref{12}).
The time domain current at any point inside the TL is obtained by performing an inverse Laplace transform, upon changing $F_{v}$ and $f_{s,v}$ by $F_{i}$ and $f_{s,v}$ in Eq.\eqref{New.1}.

\section{Radiation correction of the polarizability}
\begin{figure}[H]
     \centering
     \begin{subfigure}[b]{0.48\columnwidth}
         \centering
         \includegraphics[width=\textwidth]{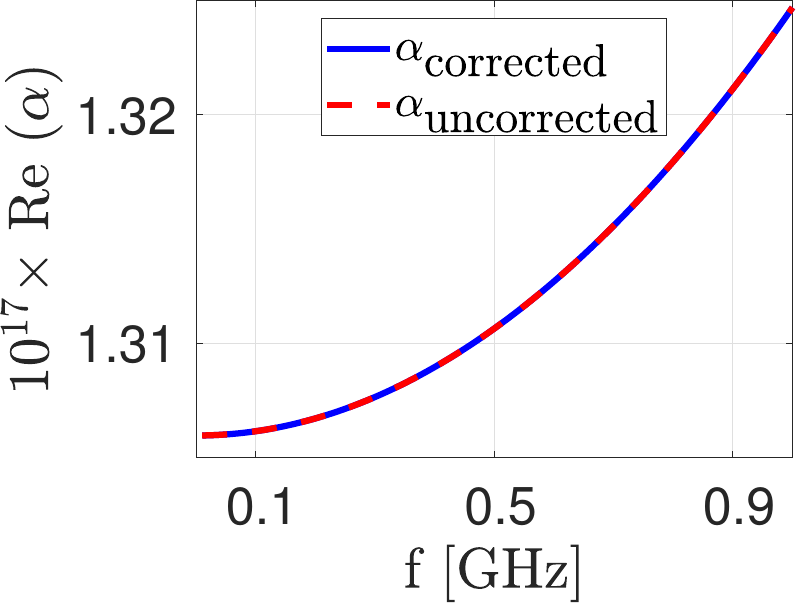}
         \caption{}
         \label{Alpha_Re}
     \end{subfigure}
     \hfill
     \begin{subfigure}[b]{0.48\columnwidth}
         \centering
         \includegraphics[width=\textwidth]{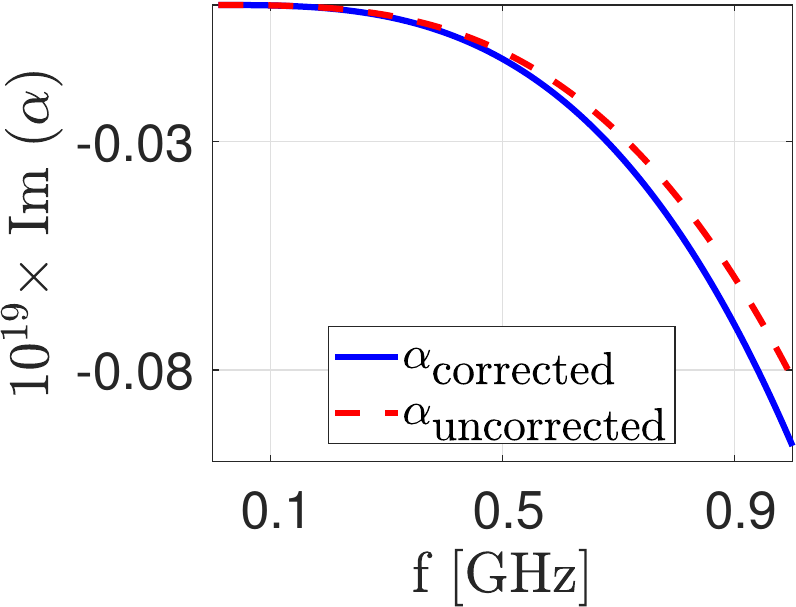}
         \caption{}
         \label{Alpha_Im}
     \end{subfigure}
\\
     \begin{subfigure}[b]{0.48\columnwidth}
         \centering
         \includegraphics[width=\textwidth]{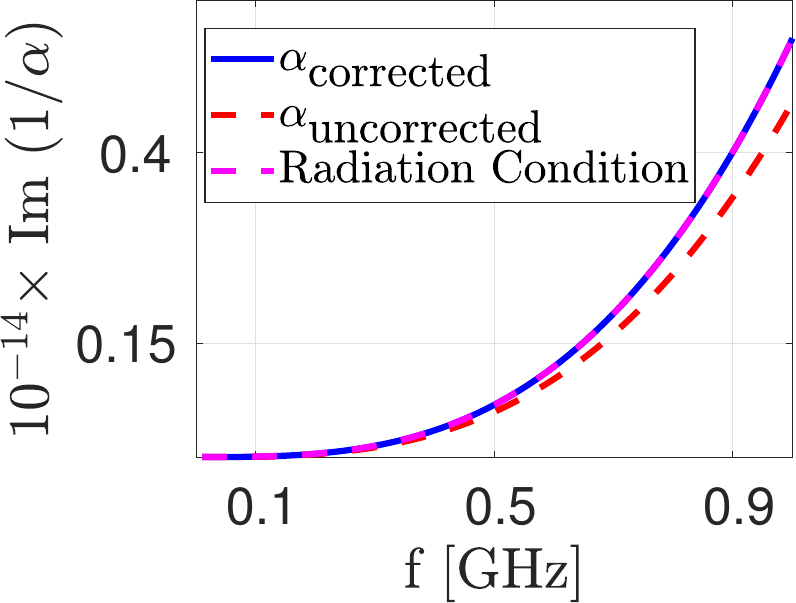}
         \caption{}
         \label{Alpha_Rad}
     \end{subfigure}
     \hfill
        \caption{Corrected and uncorrected polarizability coefficient. (a) real part, (b) imaginary part, (c) radiation condition.}
        \label{Fig_Alpha}
\end{figure}

Energy conservation implies that in the absence of material loss, the complex polarizability satisfies the radiation condition $\mbox{Im}\left(1/\alpha\right)=k^{3}/6\pi \epsilon_{0}$ \cite{tretyakov2003analytical}.
Let us denote the polarizability term discussed in section IV.2 by $\alpha_{ee}=\alpha_{\mbox{\small{uncorrected}}}$, since it does not stand under this condition. The radiation correction, $X$, is given by
\begin{equation}\label{A.1}
X=k^{3}/6\pi \epsilon_{0}-\mbox{Im}\left(1/\alpha_{\mbox{\small{uncorrected}}}\right).
\end{equation}
The corrected polarizability is defined as follows,
\begin{equation}\label{A.2}
\alpha_{\mbox{\small{corrected}}}=\frac{\alpha_{\mbox{\small{uncorrected}}}}{1+jX\alpha_{\mbox{\small{uncorrected}}}}.
\end{equation}
It is immediate to verify that $\alpha_{\mbox{\small{corrected}}}$ satisfies the radiation condition.
In Fig.~\ref{Fig_Alpha} we compare the corrected and the uncorrected coefficients. We notice that Re$(\alpha_{\mbox{\small{corrected}}})\thickapprox$ Re$(\alpha_{\mbox{\small{uncorrected}}})$ since $|$Re$(\alpha_{\mbox{\small{uncorrected}}})|>>|$Im$(\alpha_{\mbox{\small{uncorrected}}})|$ , however the imaginary part is different which compensates for the radiation condition.\\



\begin{thebibliography}{1}
\bibitem{comm2}
However, a special case occurs when $\frac{L_{1}}{C_{1}}=\frac{L_2}{C_2}$, where only a single propagating wave will be generated.
\end{thebibliography}
\end{document}